\documentclass[apj]{emulateapj}
\usepackage{graphicx}

\newcommand{\ie}{i.e.,\ }
\newcommand{\eg}{e.g.,\ }

\newcommand{\etal}{et~al.\ }
\newcommand{\ltsima}{$\; \buildrel < \over \sim \;$}
\newcommand{\simlt}{\lower.5ex\hbox{\ltsima}}
\newcommand{\gtsima}{$\; \buildrel > \over \sim \;$}
\newcommand{\simgt}{\lower.5ex\hbox{\gtsima}}

\newcommand{\magsec}{mag/arcsec$^2$}
\newcommand{\rsma}{R_{\textsc{sma}}}
\newcommand{\rsmb}{R_{\textsc{SMB}}}

\def\muv{$\mu_{\mbox{v}}$}

\def\Lsub{$L_{\rm sub}$}
\def\fsub{$f_{\rm sub}$}

\received{2010 March 16}
\accepted{2010 April 7}
\published{2010 May 5}

\begin{document}

\title{Diffuse tidal structures in the halos of Virgo
  ellipticals}

\author{Steven Janowiecki,\altaffilmark{1,2}
        J. Christopher Mihos,\altaffilmark{1} 
	Paul Harding,\altaffilmark{1}
	John J. Feldmeier,\altaffilmark{3}\break
	Craig Rudick,\altaffilmark{1} and
	Heather Morrison\altaffilmark{1}}
\email{sjanowie@astro.indiana.edu, mihos@case.edu,
        paul.harding@case.edu, jjfeldmeier@ysu.edu,
	craig.rudick@case.edu, heather@vegemite.case.edu}

\altaffiltext{1}{Department of Astronomy, Case Western Reserve University,
Cleveland, OH 44106}

\altaffiltext{2}{Department of Astronomy, Indiana University,
  Bloomington, IN 47405}

\altaffiltext{3}{Department of Physics and Astronomy, Youngstown State University,
Youngstown, OH 44555}

\begin{abstract}

We use deep V-band surface photometry of five of the brightest
elliptical galaxies in the Virgo cluster to search for diffuse tidal
streams, shells, and plumes in their outer halos ($r>50$ kpc). We fit
and subtract 
elliptical isophotal models from the galaxy images to reveal a variety
of substructure, with surface brightnesses in the range \muv = 26--29
\magsec. M49 possesses an extended, interleaved shell system
reminiscent of the radial accretion of a satellite companion, while
M89's complex system of shells and plumes suggests a more complicated
accretion history involving either multiple events or a major
merger. M87 has a set of long streamers as might be expected from
stripping of low luminosity dwarfs on radial orbits in Virgo. M86 also
displays a number of small streams indicative of stripping of dwarf
companions, but these comprise much less luminosity than those of
M87. Only M84 lacks significant tidal features. We quantify the
photometric properties of these structures, and discuss their origins
in the context of each galaxy's environment and kinematics within the
Virgo cluster.

\end{abstract}


\section{Introduction}

In a universe where structure grows hierarchically, the assembly of
galaxies and galaxy clusters is characterized by an ongoing process of
merging and accretion. These processes leave behind signatures which
can be used to learn about the dynamical history of galaxies and
clusters. On galaxy scales, structural asymmetries (\eg Zaritsky \&
Rix 1997; Conselice 2006; Coziol \& Plauchu-Frayn 2007) and diffuse
tidal streams and shells (Malin \& Carter 1980; Schweizer 1982, Tal
\etal 2009) in
field galaxies point towards recent accretion events. For galaxy
clusters, X-ray substructure (West, Jones, \& Forman 1995; Mathiesen,
Evrard, \& Mohr 1999), kinematic and spatial asymmetries in galaxy
distribution (Dressler \& Shectman 1988; Knebe \& M\"uller 2000;
Oegerle \& Hill 2001), and structure in the diffuse intracluster light
(Feldmeier \etal 2004; Mihos \etal 2005, Rudick, Mihos, \& McBride
2006) are all signatures of dynamically young events.

In field galaxies, the tidal debris formed during galaxy interactions
typically remains loosely bound to the galaxy (or merger remnant),
forming tidal tails, loops, and shells that spatially mix over many
Gyrs (Hernquist \& Spergel 1992; Hibbard \& Mihos 1995;
Gonz\'alez-Garc\'ia \& Balcells 2005). This tidal material is
typically found at very low surface brightnesses (\muv $> 26.5$
\magsec), such that very deep imaging is needed to detect it. Such
features have been used to reconstruct the dynamical history of nearby
galaxies, or determine the role mergers have played in the assembly of
specific galaxies.  For example, the tight interleaved shells seen in
some elliptical galaxies are indicative of a radial merger with a
lower mass companion (Quinn 1984), while long, luminous tidal tails
are the hallmark of major mergers (Toomre \& Toomre 1972). Fainter
tidal loops or streams are more ambiguous, and may come either from
low mass accretion events (Bullock \& Johnston 2005) or the delayed
fallback of tidal tails from a major merger (Hernquist \& Spergel
1992, Hibbard \& Mihos 1995). The variety of tidal morphologies thus
provide an important archaeological tool for understanding the
formation and growth of galaxies in the local universe.

For galaxies within clusters, the picture is more complicated. The
lifetime of diffuse tidal structure depends on two competing
processes: galaxy-galaxy and galaxy-cluster interactions which produce
tidal debris, and the dynamical heating and mixing of tidal debris in
the cluster environment as it is incorporated into the cluster-wide
ICL component (Rudick \etal 2006, 2009). The long-lived shells and loops
seen in many field ellipticals may be absent in cluster ellipticals
due to rapid tidal stripping of this loosely bound material by the
cluster potential (Mihos 2004). Alternatively, the presence of such
structure in a cluster elliptical could indicate that the galaxy is only now
being accreted into the cluster, so that cluster processes have not
yet had an opportunity to strip its diffuse tidal structure.  While
the variety of galaxy- and cluster-scale processes simultaneously at
work during cluster assembly make it difficult to unambiguously
extract the dynamical information held in tidal debris, it is clear
that the information content in these features is rich, and provides
important constraints on the dynamical history of these galaxies.

With these factors in mind, it is interesting to study the diffuse
light around elliptical galaxies in a cluster environment. As part of
our ongoing survey for ICL in the Virgo cluster (Mihos \etal 2005), we
have obtained deep, wide-field imaging of the Virgo ellipticals
M87 (NGC 4486), M86 (NGC 4406), M84 (NGC 4374), M89 (NGC 4552), and
M49 (NGC 4472). These galaxies all occupy different
environments within the cluster (see, \eg Binggeli 1999 for a review
of the structure of the Virgo Cluster). As the central dominant
elliptical in the Virgo cluster, M87 lives near the center of the
cluster potential well (as defined by the X-ray emission; Bohringer
\etal 1994), and represents the center of the most massive subgroup of
galaxies in the Virgo Cluster. M86 and M84 lie $\sim1.3$\degr\ (370
kpc)\footnote{In this work, we adopt a Virgo distance of 16 Mpc (see,
  \eg Harris \etal 1998; Ferrarese \etal 2000; Mei \etal 2007); at
this distance, 1\arcsec subtends 77.6 pc.} to the northwest of M87;
they are separated from each other by 17\arcmin\ (78 kpc) in projection, close
enough that, at faint surface brightnesses, their extended halos
appear to merge together into a common envelope of light (Mihos \etal
2005). This is likely a projection effect, however, as distance
estimates from surface brightness fluctuations place M84 about 1 Mpc
behind M86 (Mei \etal 2007). Indeed, from a combined optical and X-ray
analysis of the cluster, Schindler \etal (1999) suggest that M86 sits
at the center of its own subcluster of galaxies merging with the main
body of the cluster. M89 lies $\sim1.2$\degr\ (335 kpc) to the East of
M87 and is the least luminous of our selected elliptical galaxies.
Finally, lying 4.4\degr\ to the south of M87, M49 is the brightest
elliptical in Virgo, and defines the center of another distinct Virgo
subgroup (cluster B) which is dominated by spiral galaxies (Binggeli,
Tammann, \& Sandage 1987).  The different dynamical environments these
ellipticals find themselves in is likely to translate to differences
in the structure of their extended luminous halos.

In this work, we study the diffuse outer halos of these ellipticals,
searching for tidal structures that may trace the dynamical histories
of these galaxies. We use our deep imaging to fit and subtract a
smooth elliptical fit to each galaxy's light profile, and identify
tidal features in the residual images (for a similar approach, see,
\eg Canalizo \etal 2007). We then measure the total luminosity and
peak surface brightness of each of the cataloged features. We describe
the observational dataset and analysis techniques in \S2, and detail
the results for each galaxy in \S3.  Finally, we end with a discussion
of these features in the more general context of the hierarchical
assembly of galaxies and galaxy clusters in \S4.

\section{Imaging Data}

\subsection{Observations}

The imaging data presented here were taken as part of our ongoing
survey for diffuse intracluster light in the Virgo cluster (see, \eg
Mihos \etal 2005, 2009; Rudick \etal 2010) using Case Western Reserve
University's 0.6/0.9m Burrell Schmidt telescope located at Kitt Peak
National Observatory. The data were taken over the course of three
observing seasons in Spring 2004, 2005, and 2006. M86 and M84 were
imaged in Spring 2004, M89 in Spring 2005, M49 in Spring 2006, and M87
in both Spring 2004 and 2005 (see Table \ref{obsdata}). In all
seasons, the data were taken under dark, photometric conditions at
airmasses less than 1.5. The SITe 2048x4096 CCD images a field of view
of 0.75\degr x 1.5\degr using $1.45\arcsec$ pixels, and we build a
larger mosaic by dithering our observations by up to one degree per
field. Individual exposures are 900s long, and mosaics are made using
45--117 dithered exposures in each field. Observations are taken
through the Washington M filter, which is similar to Johnson V but
bluer by $\sim$ 300 \AA\ and cuts out the strong and variable sky
emission line at [O~I] $\lambda5577$\AA.

\begin{deluxetable}{cccc}
\tabletypesize{\scriptsize}
\tablewidth{0pt}
\tablecaption{Observational Datasets}
\tablehead{\colhead{Galaxy} & \colhead{Year} & 
\colhead{$N_{image}$} & \colhead{$N_{sky}$} }
\startdata
M87 & 2004 & 72 & 127 \\
M87 & 2005 & 45 & 103 \\
M87 & (total) & 117 & -- \\
M84/M86 & 2004 & 72 & 127 \\
M89 & 2005 & 45 & 103 \\
M49 & 2006 & 51 & 78 \\
\enddata
\tablecomments{$N_{images}$ refers to the number of images used in
  constructing each mosaic (2004 and 2005 data combined in the case of
  the M87 mosaic), while $N_{sky}$ refers to the number of blank sky 
  images used in flat fielding.}
\label{obsdata}
\end{deluxetable}

To construct accurate flat fields, we image blank sky fields
(typically 80--120 per season) at nearly the same hour angle and
declination as Virgo observations. This observing pattern reduces the
effect on the flat fields of any flexure in the telescope system.
Each night sky image had a 900s exposure time, and the brightness of
the night sky varied from 1100-1500 ADU in the images. IRAF's OBJMASK
task was used to mask stars and galaxies on the blank sky images,
after which an iterative process was applied to fit and remove planar
gradients in the sky levels before median combining the skies to
create a super sky flat (see Feldmeier \etal 2002 and Rudick \etal
2010 for complete details).

\begin{figure}[]
\centering
\includegraphics[scale=0.8]{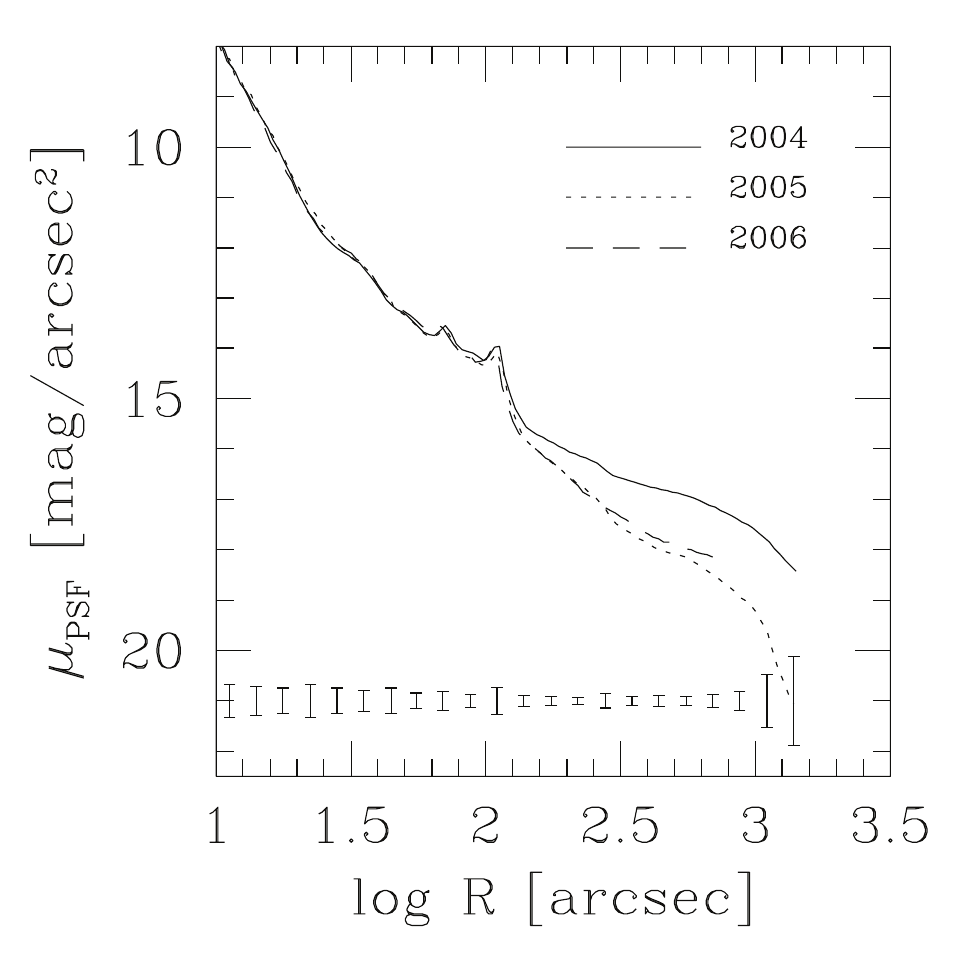}
\caption{The stellar point spread function measured in the different
  datasets. Surface
 brightness is measured relative to the central surface brightness of
 the stellar profile (\ie $\mu_{PSF,r=0}\equiv0$.) Realuminizing of the
 telescope mirror and flocking of the telescope tube between the 2004
 and 2005 seasons resulted in a significant suppression of the stellar
 wings at large radius. The double peak near log R = 2 is due to
 internal reflections within the Schmidt corrector. Errorbars along
  the bottom of the panel show characteristic uncertainties.
 } 
\label{psf}
\end{figure}

Once flattened, the images are then star-subtracted to remove the
extended wings of bright stars from the data. To construct the stellar
point spread function, we start by using short exposures of moderately
bright stars to define the small-scale PSF ($r<30\arcsec$). At larger
scales ($r>30\arcsec$) we use 900s exposures of $\alpha$Leo to
construct the PSF out to $r \sim 0.4$\degr.  The resulting large-scale
($r>10\arcsec$) PSFs are shown in Figure \ref{psf}. Between the 2004
and 2005 seasons, the primary mirror was realuminized and the interior
of the telescope tube was flocked with black velvet to reduce
scattered light. These improvements reduced the brightness of stellar
wings by about 2.5 mag/arcsec$^2$ at large radius ($r\sim
20\arcmin$). With the reconstructed PSF, we then mask stars out to the
radius where their scaled PSF falls below 3 ADU (\muv $\approx 28$)
and then subtract the extended wings out to 0.5 ADU (\muv $\approx
30$).  For the 2004 data reduction, star subtraction was done on the
original images before sky subtraction and registration; for the 2005
and 2006 datasets, star subtraction was done after sky subtraction and
registration. For the M87 dataset, which encompasses data taken both
in 2004 and 2005, we have re-reduced the 2004 images for consistency;
however, this test showed that differences between the two methods are
small as long as sufficient area is available far away from bright
stars to set the sky background.

Sky subtraction is problematic because the size of the Virgo Cluster is
much bigger than the imaging field of view. First, we mask the images
of all sources brighter than 2.5$\sigma$ above sky, then spatially
rebin the images by calculating the mode of the remaining unmasked
pixels in 32x32 pixel ``superpixels.'' Regions of images far away from
bright stars and galaxies are then selected in which to measure a
preliminary sky value to subtract. At this point, the binned images
are registered and an iterative plane fitting process is employed to
fit residual sky planes to the binned images, constrained to minimize
frame-to-frame variations in spatially-overlapping regions. The
resulting edge-to-edge sky gradients which are removed from the images
are typically about 5 ADU, or $<$0.5\% of the absolute sky level
(1100-1500 ADU, or \muv=21.6-21.3).  Once sky subtraction is complete,
we register and median combine the individual images to create the
final mosaic.

We quantify photometric uncertainty on two scales: small scale
uncertainty due to read noise, photon noise, and small-scale
flatfielding errors, and large scale uncertainties due to uncertainty
in large scale flat fielding and sky subtraction. Small-scale
uncertainties can be ``root-N reduced'' by combining many individual
images together into the final mosaic, and also by spatially binning
the mosaic. Large-scale uncertainties do not reduce so easily, and we
characterize them as a floor to our uncertainty, as described
below. We refer the reader to Morrison \etal (1997), Feldmeier \etal
(2002), Mihos \etal (2009), and Rudick \etal (2010) for more details
of our error model as applied to the deep surface photometry.

\begin{figure*}[]
\centering
\includegraphics[width=6in]{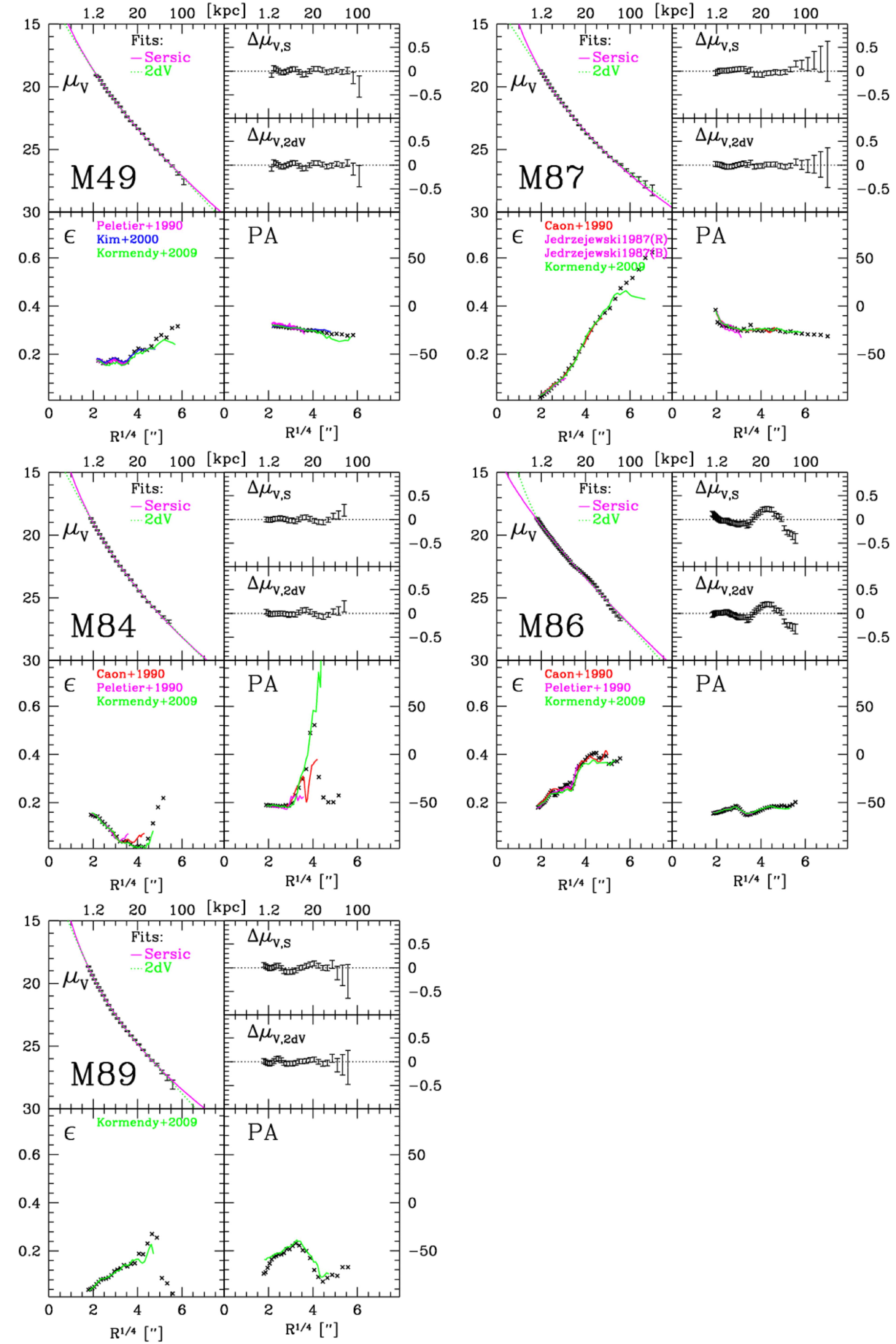}
\caption{
Isophotal fits for each of the five galaxies. The legend for each 4x4
panel is as follows:
Upper left: Surface brightness profile, along with the best fit
S\'ersic (solid magenta line) and double deVaucouleurs (dotted green
line) profiles. 
Upper right: Residuals from S\'ersic ($\Delta \mu_{V,S})$, and
double deVaucouleurs ($\Delta\mu_{V,2dV}$) surface brightness
profile fits.
Lower left: Ellipticity ($\epsilon$) as a function of radius,
with profiles from the literature over-plotted as noted.
Lower right: Position angle (PA) as a function of radius, 
with profiles from the literature over-plotted as noted.
}
\label{allfits}
\end{figure*}

For the CCD used in these observations, the gain was 2e$^-$/ADU and
readnoise was 12e$^-$ or 6 ADU. The typical sky background was 1250
ADU, which produces photon noise of 24.5 ADU per pixel. Photon noise
in the blank skies used to construct the flat field will also result
in small-scale flatfielding uncertainty (which will be mediated
significantly by our dithering of object images). The resulting
small-scale flat fielding error is 2.7 ADU (see Mihos \etal
2008). These errors add in quadrature to give an uncertainty of 25
ADU/pixel in a single Virgo image. When median combining $N$
individual images together, the noise goes down as $1.22\sqrt{N}$;
for a typical dataset analyzed here, $N \approx 30$, giving a per
pixel noise in the final mosaic of 0.6 ADU. Furthermore, when
examining residual images, we spatially rebin the images in blocks of
9x9 pixels; these rebinned images then have a per pixel uncertainty of
$<$0.1 ADU due to these small-scale effects.

Over larger scales, however, the dominant uncertainty lies in the
large-scale flat fielding and sky subtraction. Flat fielding errors
can come from a variety of imperfections in the blank sky images used
to construct the flat, such as spatial variations in the sky
brightness, the unsubtracted extended wings of bright stars, and
diffuse low surface brightness structures such as galactic
cirrus. Night sky variations in the object images also lead to
uncertainty in the sky subtraction, as does the lack of substantial
areas of pure sky in which to define the sky levels. Variability in
the night sky happens both randomly and systematically (with, for
example, hour angle, time of night, or airmass). To quantify these
combined uncertainties, we measure the residual flux in 100x100 pixel
boxes in blank sky regions in the flattened, sky-subtracted
mosaic. These boxes show a mean residual sky background of 0.1 +/- 0.5
ADU, and so we adopt a conservative 1 ADU uncertainty as the floor to our error
model. Given our photometric solution, 1 ADU corresponds to a
surface brightness of \muv=29.2.

\subsection{Ellipse Fitting and Model Subtraction}

Our approach to removing the smooth light profile from these galaxies
involves fitting nested elliptical annuli over the galaxy's entire
extent, and then subtracting the best-fit model from the image.  As
this fitting process is affected by all of the light in each isophote,
it is imperative to remove any extra sources of non-galaxy light, lest
they be interpreted as parts of the galaxy. The star mask and
subtraction from the image reduction eliminates most of the light from
the stars in each field, but other sources still remain. We have
masked, by hand, all light in each image from nearby galaxies as well
as the bright wings, diffraction spikes, and column bleeds around
bright stars. To mask all small-scale sources of light, we use an
automated source detection algorithm on an image with large scale
structures removed. The large-scale structure is removed by smoothing
the original image with a ring-median annulus of inner and outer radii
of $4''$ and $12''$ ($300$pc and $900$pc), and subtracting it from the
original image. IRAF's OBJMASKS task is used to identify all sources
$2\sigma$ above sky which cover at least $2$ pixels. This mask is
combined with the previous mask to obtain our total mask. However, the
galaxy centers are almost completely covered by this total mask. To
obtain a reliable surface brightness profile fit, we unmask the
central $2.5'$ of each galaxy before fitting.

We use IRAF's ELLIPSE task to fit elliptical isophotes to each galaxy,
which implements the iterative technique described by Jedrzejewski
(1987).  At each step the geometric parameters of the elliptical
isophote are varied until the residuals are minimized. For most of the
galaxies we use a fixed center for the isophotes, which we determine
by manually varying it until the inner $\sim 0.5'$ shows no asymmetric
residuals in the model-subtracted images.  Within each isophote, the
following parameters are computed: mean intensity, ellipticity ($1 -
\rsma/\rsmb$) and position angle (measured north through east). The
fitting process only includes 1st and 2nd order Fourier terms
(conveyed in the ellipticity and position angle) and does not fit the
3rd or 4th order terms. The 4th order ($A_4$) component is a sign of
disky or boxy isophotes (see Kormendy \etal 2009 and references
therein), which will manifest in our subtracted images as a
butterfly-like pattern of alternating light and dark wedges in the
inner regions ($r \lesssim 0.5-1 r_e$) of the galaxies. When later
analyzing substructure, we only consider substructures outside of
these features. We also avoid fitting the inner regions of the
galaxies as they are often saturated and also our large pixel scale
($1.45\arcsec$/pixel) is unable to capture the strong gradient
around the center. In the low surface brightness outskirts of the
galaxies, the robustness of the fit is determined by uncertainties in
the sky subtraction. We continue fitting isophotes until we reach an
isophotal surface brightness which is indistinguishable from sky. The estimated sky
uncertainty in our images is $\sim 1$ ADU; to determine the effects
that this has on our fits, we carry out the same fitting procedure on
an image which has $1$ ADU added to every pixel and on another with
$1$ ADU subtracted. These two extra fits are how we obtain the error
bars which are displayed on the best fit profiles.

Finally, we use IRAF's BMODEL task to make a model galaxy from the
best fit elliptical isophotes. We subtract this model galaxy from the
initial image and examine the residuals. The residuals contain both
bright and dark regions, where the best-fit model under- and
over-subtracts the light, respectively. Initially, more artifacts were
evident in these residuals, around the edges of our hand mask, but we
have iteratively adjusted our mask until all of the light coming from
the other stars and galaxies is masked. We tried running the object
detecting algorithm again on these residuals to improve the total
mask, but found that the improvements in the fit and subtraction were
minimal. This final subtraction is median smoothed into $7''$x$7''$
bins, to bring out faint structures.

\subsection{Surface Brightness Profile Fits}

As a consistency check, we also fit the extracted surface brightness
profiles to both S\'ersic and double-de Vaucouleur profiles, and
compare to values in the literature; the fitted parameters and their
1$\sigma$ uncertainties are reported in Table \ref{sbfits}. In the
fitting process, we take a rather conservative error model, wherein
each datapoint is assigned an uncertainty equal to the maximum of 0.05
\magsec or 1 ADU, whichever is larger.  The former value corresponds
to uncertainty in the photometric zeropoint and isophotal fitting
process at high surface brightness, while the latter value refers to
sky subtraction uncertainty, which dominates at low surface
brightness. This error model leads to reduced $\chi^2$ values for the
fits which are less than one, suggesting that this error model is a
conservative one, overestimating the uncertainties in the fit. Since
we are mostly interested in demonstrating consistency with other
values reported in the literature, we do not further refine the error
model. We have verified, however, that reasonable changes in the
adopted uncertainties do not lead to large systematic differences in
the extracted profile parameters.

Kormendy \etal (2009; hereafter K09) have recently published an
extensive set of surface photometry for Virgo ellipticals, with
detailed S\'ersic fits. 
Our S\'ersic fits are in reasonably good
agreement with those of K09, with the fitted parameters typically
agreeing to within 1-2$\sigma$. Where there is noticeable
disagreement, the discrepancy can generally be attributed to the
different radial ranges being fit between the two studies; as shown in
K09, the fitted parameters can be sensitive to this choice.\footnote{We note
that our quoted uncertainties in Table\label{sbfits} are standard
$1\sigma$ uncertainties, but direct the reader towards the more
complete discussion of parameter uncertainties in S\'ersic fits in K09.}
In surface
photometry of more distant clusters, Gonzalez \etal (2005) have
proposed that brightest cluster galaxies (BCGs) are better fit by
double-deVaucouleur (hereafter ``2dV'') profiles, and that the
properties of the outer profile are indicative of a diffuse
intracluster component. Alternatively, Seigar \etal (2007) argue that
S\'ersic + exponential profiles are better able to fit the extended
light from cluster cD galaxies. Our sample, of course, is not a sample
of BCGs, but rather a sample of luminous ellipticals {\it within} an
individual cluster, only one of which can be a BCG. Nonetheless, for
comparison with these other studies, we fit our profiles to both
S\'ersic and double-deVaucouleur models. In most cases, we find either
fit acceptable, and therefore do not attempt fits which have
additional fitting parameters. While the evidence for 2dV profiles as
a superior function over single S\'ersic is marginal at best in our
dataset, we do calculate the relative amount of light in the inner and
outer components of the 2dV, for comparison with the results of
Gonzalez \etal (2005).

\begin{deluxetable*}{ccccccccccc}
\tabletypesize{\scriptsize}
\tablewidth{0pt}
\tabletypesize{\tiny}
\tablecaption{Analytic Surface Brightness Profile Fits}
\tablewidth{0pt}
\tablehead{
\colhead{} & \multicolumn{4}{c}{S\'{e}rsic fit} & 
\colhead{} & \multicolumn{5}{c}{Double deVaucouleur  fit}\\
\cline{2-5} \cline {7-11}\\
\colhead{Galaxy} & \colhead{$n$} & \colhead{$r_e$} & 
\colhead{$\mu_e$} & \colhead{$\chi^2$} &  
\colhead{Galaxy} &
\colhead{$r_e$ (in)} & 
\colhead{$\mu_e$ (in)} & \colhead{$r_e$ (out)} & 
\colhead{$\mu_e$ (out)} & \colhead{$\chi^2$}\\
\colhead{} & \colhead{} & \colhead{[arcsec]} & \colhead{[\magsec]} & \colhead{} & 
\colhead{} & \colhead{[arcsec]} & \colhead{[\magsec]} &
\colhead{[arcsec]} & \colhead{[\magsec]} &
\colhead{}
}

\startdata
 M49 & 6.90 (0.46) & 311 (20) & 23.8 (0.29) & 0.80 &
 M49 & 52.4 (20.2) & 21.2 (0.5) & 423 (86) & 24.5 (0.6) & 0.75\\
 M87 & 11.0 (0.9) & 666 (101) & 25.6 (0.7) & 0.91 &
 M87 & 81.5 (8.3) & 21.7 (0.1) & 1753 (440) & 27.5 (0.5) & 0.26\\
 M86 & 5.18 (0.17) & 372 (17) & 24.2 (0.15) & 6.1 &
 M86 & 6.1 (2.7) & 18.4 (3.2) & 335 (10) & 23.9 (0.1) & 4.3 \\
 M84 & 9.75 (0.85) & 204 (18) & 23.8 (0.5) & 0.55 &
 M84 & 40.4 (6.9) & 20.9 (0.25) & 501 (128) & 25.9 (0.6) & 0.61\\
 M89 & 14.6 (1.9) & 321 (69) & 25.3 (2.2) & 1.1 &
 M89 & 15.9 (2.6) & 19.8 (0.3) & 324 (39) & 25.3 (0.3) & 0.51\\
\enddata
\tablecomments{Uncertainties in the fitted parameters are
 given in parentheses.}
\label{sbfits}
\end{deluxetable*}

\subsection{Photometric Measures of Substructure}

We employ a few methods to quantify the nature of the features visible
in the residual images of the model subtraction. First, we identify
individual features located outside the central regions ($r \lesssim
0.5-1 r_e$) of the galaxies where higher order Fourier terms usually
dominate the residual image and away from any other galaxies in the
frame. To ensure that the features we identify are not artifacts of
the subtraction process, we verify in all cases that they are present
(although much harder to see) in the unsubtracted images as well.
Once confirmed, we calculate the luminosity of these discrete features
in polygonal apertures containing the features, and using adjoining
regions to determine the local background for accurate background
subtraction.  The subtraction of a locally-defined background from the
luminosity of each feature ensures that small-scale background
uncertainty due to spatial variation of the subtracted galaxy model is
minimized. Even so, the background subtraction remains the biggest
source of uncertainty in measuring the flux in each region. We
quantify this error by making small adjustments to the background
regions to quantify the resulting background flux (see Rudick \etal
2010 for more details). We find this variation to be $\sim \pm$ 0.25
ADU, which we propagate to uncertainties in the derived luminosities.

As an independent check of the robustness of the derived substructure
luminosities, we also reduced a completely independent dataset imaging
M49 in 2005, and compared it to the results for the 2006 M49 dataset
described here. We find the derived luminosities of the features agree
to $\sim$ 5-10\%.

In addition to calculating the total luminosity of each feature, we
also compute the peak surface brightness. We quantify this by
constructing a $7\arcsec$x$7\arcsec$ (0.5x0.5 kpc) median smoothed,
model subtracted image, and examining the pixel intensities in each
feature. To avoid statistical fluctuations associated with identifying
the brightest binned pixel, we define peak surface brightness to be
the 90th percentile of the intensity distribution. A visual
examination of the images shows this to be robust against
contamination from unsubtracted background objects and stars.

\section{Results for Individual Galaxies}

\subsection{M49 (NGC 4472)}

We begin with brightest Virgo elliptical, M49, located in the Virgo
Southern Extension, four degrees south of M87.  The final mask for M49
includes all of the bright stars in the field, as well as nearby faint
galaxies (e.g. NGC4492, NGC4488). We fit the isophotal model using
IRAF's ELLIPSE task, as described in \S2.2, using fixed center (at
$\alpha =$ 12:29:46.8, $\delta =$ +8:00:01.8 J2000).  Our extracted
surface brightness profile never reaches our limiting magnitude of
\muv$ = 29$; the fit stops converging past $\rsma = 32.8'$ (\muv$ =
27.5$) where too much of the galaxy extends off the edges of our
image. The best fit profiles for surface brightness, position angle,
and ellipticity are shown in Figure \ref{allfits}. We compare our
results to Peletier \etal (1990) with CCD surface photometry in $BR$
bands, Kim \etal (2000) who used a CCD in Washington $CT_{1}$ bands,
and the composite $V$ band profile of K09.  As can
be see in Figure \ref{allfits}, the ellipticity and position angle
profiles are in very good agreement between these studies.

The isophotal fits to S\'ersic and 2dV profiles are shown in Figure
\ref{allfits} and Table \ref{sbfits}. For the S\'ersic fit our
extracted parameters agree reasonably well (within $2\sigma$) with
those measured by K09. There is no significant difference in the
quality of fit between the S\'ersic and 2dV models; extracting a total
luminosity for the galaxies from the fitted profiles yields $L_V = 1.6
{\rm\ and\ } 1.5\times 10^{11} L_\sun$ for the S\'ersic and 2dV fits,
respectively. In the 2dV fit, the outer component carries 50\% of the
total galaxy luminosity.

\begin{figure*}[]
\centering
\includegraphics[width=6.25in]{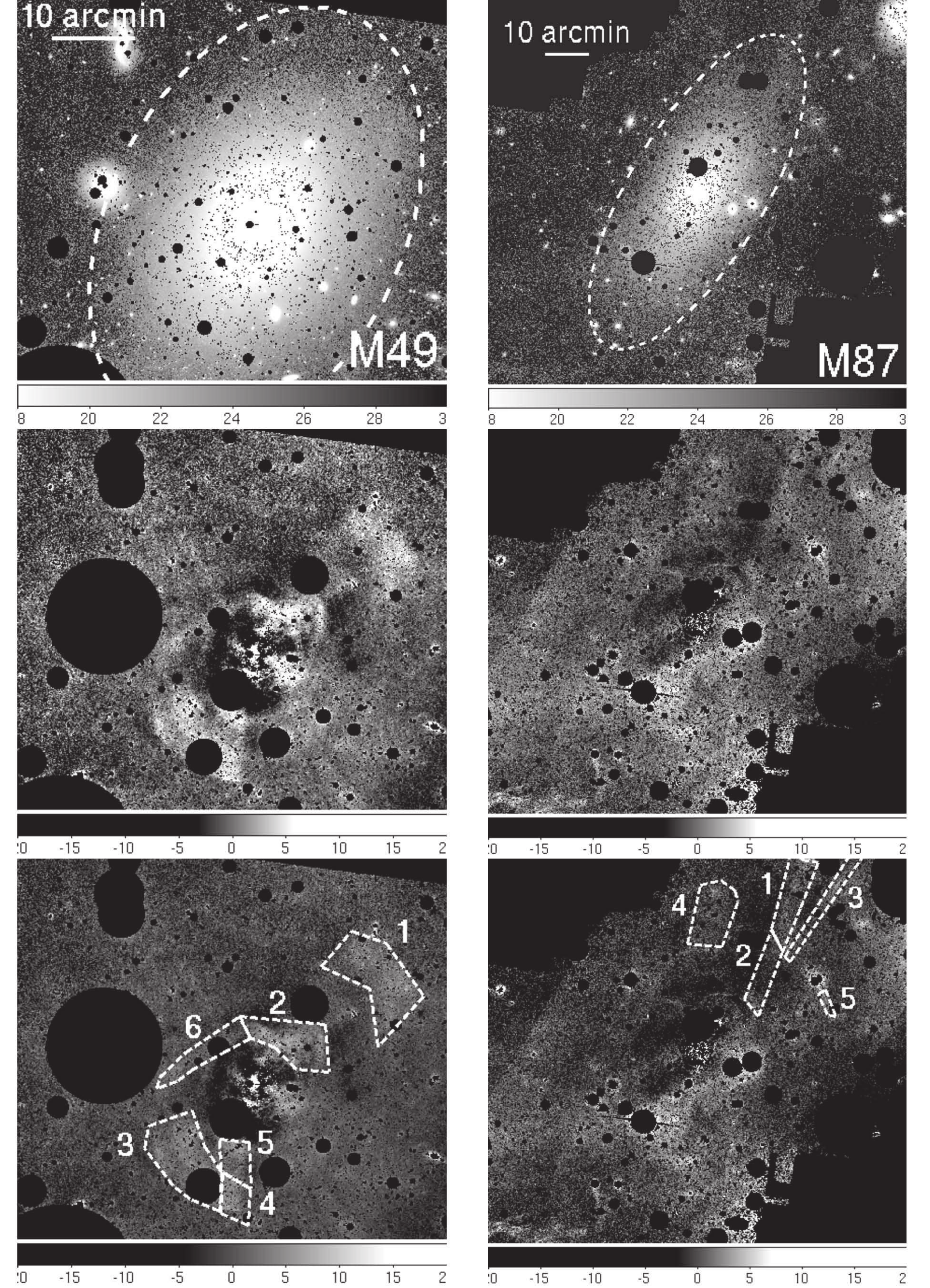}
\caption{
Deep imaging for M49 (left column) and M87 (right column). In each
column, the top panel shows the original image, masked using the
star mask from the data reduction. The greyscale legend shows surface
brightness in V \magsec, and the overlaid ellipse outlines the 
\muv=29 \magsec\ isophote. The middle panel shows the model subtracted
residual image, median smoothed on 5x5 pixel (7\arcsec x 7\arcsec)
boxes. In this panel, the mask shown is the final mask used in
the fitting and subtraction process, and the greyscale legend shows
brightness in ADU, where 1 ADU corresponds to \muv=29.2 \magsec. 
The butterfly-like pattern near the center of M49 is caused by high
order (A4) shape terms which are not captured in our ellipse
fitting. We ignore these regions when searching for residual
substructure. Similarly, we do not identify substructure in the 
southeastern (lower left) quadrant of the M87 image
where there is significant contamination from galactic cirrus (see Rudick \etal
2010). The bottom panels show the same image as the middle
panel, but with region identifier boxes overlaid. 
}
\label{subtract_m49m87}
\end{figure*}

We subtract the extracted profile from the original image of M49 to
yield our residual image, shown in Figure
\ref{subtract_m49m87}. Strikingly visible in the residual image is a
complex system of diffuse shells. A large shell sits 19.5\arcmin\ (90
kpc) to the northwest (Region 1), while a double-edged shell (Region
3) sits on the opposite side of M49, 12.5\arcmin\ (60 kpc) to the
southeast. There is also a complex inner structure (Region
2) 7\arcmin\ (30 kpc) to the northwest of M49, which appears to be the
overlapping of several discrete shells. We also see a plume south of
M49 center (Region 4), which partially overlaps the SE shell (Region
5). Finally, we identify a faint plume to the northeast (Region 6),
running through the galaxy VCC 1254, and perhaps connecting with the
inner shell structure. The peak surface brightnesses and total luminosities of these
features are given in Table \ref{m49tab}. While these shells were not
seen in the recent Tal \etal (2009) study of tidal features around nearby
ellipticals, it is likely due to the extended nature of the shells;
the field of view of the Tal \etal images was too small to reach most
of these outer features. We also note at even fainter surface
brightnesses a hint of a plume running to the northeast of M49, but
this (unmarked) plume is only marginally detected, and we do not
photometer it. 

We note in the context of these shells that if the isophotal model has
spurious features in it (due to rapidly changing ellipticity and
position angle in the derived ELLIPSE model), the subtraction process
can imprint shell-like artifacts into the residual image. We have been
very careful in the fitting process to avoid these types of
features. The fitted isophotal parameters vary smoothly with radius,
and we have checked that the best-fit model did not have any structure
that would be imprinted on the residual image, and we have also
verified that all these structures can be recovered in the
unsubtracted images.

\begin{deluxetable}{cccc}
\tabletypesize{\scriptsize}
\tablewidth{0pt}
\tablecaption{M49 Substructure}
\tablehead{\colhead{Region} & \colhead{Name} & 
\colhead{$\mu_{V,\textrm{peak}}$} & \colhead{L$_{\textrm{tot},V}$} \\ 
\colhead{\#} & \colhead{} & \colhead{[\magsec]} &
\colhead{[$10^8$ $L_\odot$]} } 
\startdata
1 & NW Shell       & 27.6 & 2.28 (0.18) \\
2 & Inner Shells   & 27.0 & 2.03 (0.11) \\
3 & SE Shell       & 27.4 & 1.57 (0.13) \\
4 & Overlap of 3\&5& 27.4 & 0.60 (0.04) \\
5 & S Plume        & 27.0 & 0.51 (0.04) \\
6 & NE Plume       & 27.7 & 0.30 (0.04) \\
\cline{2-4}
 & Total          &      & 7.29 (0.25)
\enddata
\tablecomments{Uncertainties on luminosities are given in parentheses.}
\label{m49tab}
\end{deluxetable}

The S Plume (Regions 4 \& 5) also lies very near the tidally disturbed
dwarf irregular UGC 7636 (in our image, UGC 7636 itself lies under the
circular mask immediately north of the S Plume). Patterson \& Thuan
(1992) studied UGC 7636 using optical and 21-cm imaging, and found
short optical tidal tails extending $\sim$ 1\arcmin\ north and south of
the dwarf, and evidence for ram pressure stripping of its neutral
hydrogen gas. The linear, low surface brightness S Plume is aligned
with the relatively high surface brightness tails identified by
Patterson \& Thuan, and is likely an extension of the southern tidal
tail to larger radius. The narrowness of the plume argues, however,
that this interaction is not the same that gave rise to the much
broader, more extended NW and SE Shells.

The interleaved structure of those shells, roughly along the major axis
of the M49, is very reminiscent of simulations of the accretion and
subsequent disruption of small satellite galaxies around larger
ellipticals (Quinn 1984, Hernquist \& Quinn 1988, 1989). While major
mergers can also leave behind shell-like structures (\eg Hernquist \&
Spergel 1992), these are generally not so aligned and interleaved the
way the shells in M49 are. These shells are also extremely sharp: a
radial cut along the NW Shell (Region 1) shows a sudden 2 \magsec\ drop
over 35\arcsec\ (2.7 kpc), while the SE Shell (Region 3) shows a steep
1.5 \magsec\ drop over 50\arcsec\ (3.8 kpc), while the innermost shell
(Region 2) has a much more diffuse decline in radial intensity. The
fact that the sharpness of the shell is correlated with its distance
from the center of M49 is consistent with an accretion origin, where
the inner shells have experienced several dynamical crossings, resulting
in heightened orbital mixing and diffusion of the shell's
sharpness. The sharpness of the outer shell argues that it has been
relatively unperturbed during its lifetime; we return to this
implication in \S4.

\subsection{M87 (NGC 4486)}

The second galaxy we study is M87, the giant elliptical at the
heart of the Virgo Cluster.  Its final mask includes all of the small
galaxies nearby (notably NGC 4478, 4476) and several diffraction
spikes and saturation bleeds from bright stars. Our fit uses a fixed
center (at $\alpha=$12:30:49.4, $\delta=$+12:23:28.2 J2000), and
extends outward to $\rsma \sim 40\arcmin\ $(180 kpc), where the surface
brightness drops below \muv $=29$ \magsec, our 1$\sigma$ photometric
limit. For display purposes, our subtracted images show residuals from
an extended fit which continues into this sky level, but our
quantitative discussion and profile fitting are restricted to the part
of the profile with \muv $<29$.  Figure \ref{allfits} shows
M87's surface brightness profile and other best fit isophote parameters.

Also plotted in Figure \ref{allfits} with our ellipticity and position
angle profiles are results from similar previous studies of M87.
Jedrzejewski (1987) published surface photometry of M87 in both B and
R, while Caon \etal (1990) studied M87 using a combination of B-band
CCD imaging and wide-field Schmidt photographic plates. K09 also
combine their own observations with surface photometry published in
the literature to construct a composite surface brightness profile of
M87 in the V-band.  While our data extend to somewhat larger
semi-major axis, there is, for the most part, good agreement in the
geometric isophotal parameters in regions of overlap. At large radius
($R^{1\over 4}> 5.5$, or $R>15$\arcmin) there is some divergence in
the ellipticity, but this is in regions of very low surface brightness
(\muv$>$27.5) and may be due to contamination from galactic cirrus to
the southeast of M87 (see discussion below).

The isophotal fits to S\'ersic and 2dV profiles are shown in Figure
\ref{allfits} and Table \ref{sbfits}. For the S\'ersic fit, our
extracted parameters agree well with those determined by K09. We note
that 2dV is a somewhat better fit to the profile than the S\'{e}rsic
model; however, much of the discrepancy between the fits comes in the
outer regions at very low surface brightness. The total luminosity of
M87 is $L_V = 1.1 {\rm\ and\ } 1.2\times 10^{11} L_\sun$ for the
S\'ersic and 2dV fits, respectively, and in the 2dV model the outer
component accounts for 50\% of the total luminosity.

We subtract the smooth isophotal model for M87 from the original image
and show the subtracted image below the original in Figure
\ref{subtract_m49m87}. A major source of uncertainty is immediately
visible in this image -- contamination by back-scattered Galactic light
from the galactic cirrus (Sandage 1976; Witt \etal 2008; Rudick \etal
2010). Much of the residual structure seen to the
southeast (lower left) of M87 correlates strongly with far infrared maps of
Galactic dust (Schlegel \etal 1998; Miville-Desch\^{e}nes \& Lagache
2005), suggesting much of what we are seeing is due to scattering from
galactic dust. It is particularly troublesome that
this emission connects smoothly with the broad excess of light to the
southeast of M87 which Weil \etal (1997) attributed to tidal encounter; this
feature may be nothing more than scattered light from galactic cirrus.
Given the likelihood of contamination by dust, we avoid attributing
{\it any} substructure in this region to true tidal features around M87
itself. We limit our analysis to regions to the northwest of M87 which are
largely free of dust contamination.

Highlighted in Figure \ref{subtract_m49m87} are several extended
features surrounding M87. There is a wide plume extending radially to
the north of M87's center (Region 4), and also two narrow streams
extending to the northwest (Regions 1+2 and 3).\footnote{Regions 1 and
2 make up a single stream. Region 1 refers to the portion of the
stream which extends beyond M87's halo, while Region 2 refers to the
portion inside the halo. We photometer the stream separately in these
two pieces, so that appropriate backgrounds can be applied.} These
radial streams are bright enough that they are easily visible even in
the unsubtracted deep image (Mihos \etal 2005). The nearby
galaxy PGC 41098 (VCC1148) also has a small stream emanating from it
(Region 5). There is a hint that this stream continues to the
northeast, running through the radial stream, and connecting up with
the N Plume (Region 4), but this is just at our surface brightness limit
we do not consider it firmly enough detected to photometer. The
luminosity and maximum surface brightness of the detected features are
given in Table \ref{m87tab}.

To assess the authenticity of all these features, and avoid confusion with
galactic cirrus, we have compared them to far infrared IRIS
observations over the same area (Miville-Desch\^{e}nes \& Lagache
2005). While the spatial resolution of the IRIS data is only
4.3\arcmin, we find no correlation with regions of suspected galactic
dust contamination, and are confident that these features are true
stellar features around M87. Again, however, since we are explicitly
avoiding the dust-contaminated regions to the southeast of M87, our catalog
of features is likely underestimating the total structure around M87.

The long linear streams to the northwest of M87 are suggestive of small
satellites falling in on radial orbits, or on more tangential orbits
viewed along the orbital plane. The larger NW Stream (Regions 1+2)
crosses the galaxy pair NGC 4458/61. These two galaxies have a
velocity difference of 1300 km/s and are not likely engaged in any
slow mutual interaction that would draw out long tidal tails. It is
possible, however, that the stripping of one of these galaxies as it
orbits in the potential well of the cluster could have given rise to
the NW Stream. The thinner WNW Stream (Region 3), projects across the
dwarf galaxy VCC 1149, and again could be due to stripping of this
galaxy as it orbits M87.

\begin{deluxetable}{cccc}
\tabletypesize{\scriptsize}
\tablewidth{0pt}
\tablecaption{M87 Substructure}
\tablehead{\colhead{Region} & \colhead{Name} & 
\colhead{$\mu_{V,\textrm{peak}}$} & \colhead{L$_{\textrm{tot},V}$} \\ 
\colhead{\#} & \colhead{} &  \colhead{[\magsec]} &
\colhead{[$10^8 L_\odot$]} } 
\startdata
1+2 &  NW Stream (N+S)    &  27.8  &  2.35 (0.28) \\
3   &  WNW Stream         &  28.3  &  0.53 (0.18) \\
4   &  N Plume            &  28.1  &  1.41 (0.31) \\
5   &  W Arc              &  28.2  &  0.13 (0.02) \\
\cline{2-4}
   &  Total               &        &  4.42 (0.46)
\enddata
\tablecomments{Uncertainties on luminosities are given in parentheses.}
\label{m87tab}
\end{deluxetable}

\subsection{M84/M86 (NGC 4374/4406)}

The surface brightness profiles of M84 and M86 overlap at large radii
(where \muv $\ga 25$) and thus cannot be fit
independently. We have employed an iterative process, by alternately
fitting and subtracting both galaxies. Our fits show that M84 is
better fit with a fixed center ($\alpha=$12:25:03.8,
$\delta=$+12:53:13.2 J2000) and that M86's best fit isophotes have a
center that drifts south east about $150$ arcseconds from the initial
center of ($\alpha=$12:26:11.8, $\delta=$+12:56:46.6 J2000). The
centroid drift is small at high surface brightness, but is more
significant for the faint outer isophotes. At \muv = 25, the
center has drifted less than $10''$, and the more significant drifting
occurs fainter than \muv = 27.

\begin{figure*}[]
\centering
\includegraphics[width=6.25in]{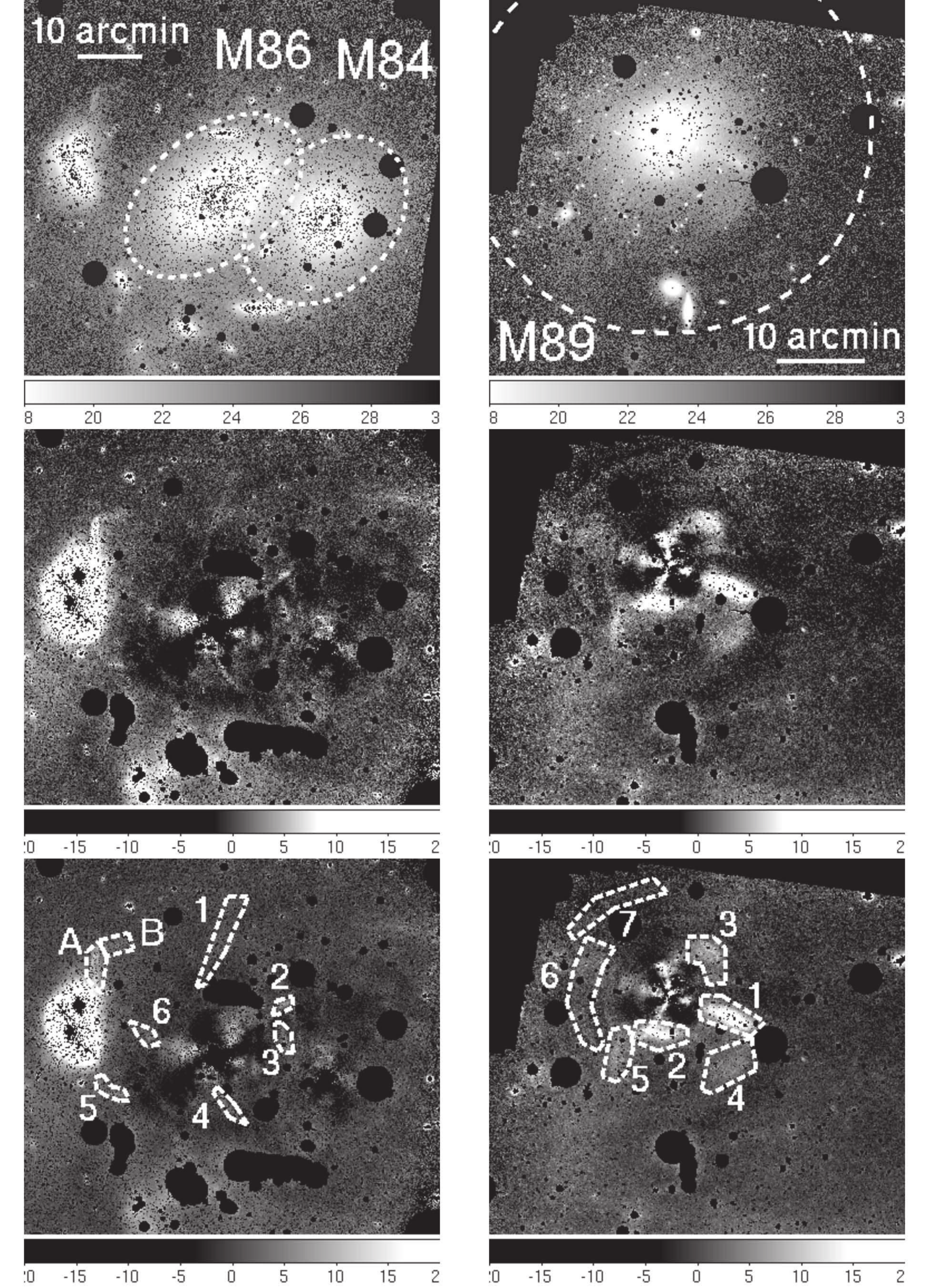}
\caption { 
Deep imaging for M86/M84 (left column) and M89 (right column). In each
column, the top panel shows the original image, masked using the
star mask from the data reduction. The greyscale legend shows surface
brightness in V \magsec, and the overlaid ellipse outlines the 
\muv=27 \magsec\  (for M86/M84) and \muv=29 \magsec\ (for M89) isophotes.
The middle panel shows the model subtracted
residual image, median smoothed on 5x5 pixel (7\arcsec x 7\arcsec)
boxes. In this panel, the mask shown is the final mask used in
the final photometering of the substructure, and the greyscale legend shows
brightness in ADU, where 1 ADU corresponds to \muv=29.2 \magsec. 
The butterfly-like pattern near the center of M49 is caused by high
order (A4) shape terms which are not captured in our ELLIPSE
fitting. We ignore these regions when searching for residual
substructure. The bottom panels show the same image as the middle
panel, but with region identifier boxes overlaid. 
}
\label{subtract_m84m86m89}
\end{figure*}

We begin our iterative process by masking M86 and making an initial
fit to M84, with limited radial extent. We subtract this M84 fit from
the image, mask the residuals near M84's center, unmask M86, and make
an initial fit to M86. We then subtract the M86 fit from the original
image, again masking the inner residual, and make a new fit to M84,
over a larger radial range. We continue this iterative process for 5
steps, until we have fit both galaxies out to $\rsma =$ 34\arcmin\ for
M86 and $\rsma =$ 24\arcmin\ for M84.  Because of the complexity of the
fitting process, and the crowded nature of the field surrounding M84
and M86, our fits do not extend to the nominal limit of \muv = 29. As
noted above, we find that as we reach \muv = 27, the centroid of the
fit begins drifting significantly, at about the same point where the
isophotes begin encompassing other galaxies in the field. We therefore
take this brighter limit of \muv = 27 as the limit of our fitting
process when extracting the analytic profile fits, and show this as
our outermost isophote in Figure \ref{subtract_m84m86m89}.  The
isophotal ELLIPSE fits for both M84 and M86 are shown in Figure
\ref{allfits}.

We compare our surface brightness, ellipticity, and position angle
profiles with those of Caon \etal (1990), Peletier \etal (1990), and
K09 in $B$, $R$, and $V$ bands, respectively.  For M84, we again find
good agreement between those studies and ours, save for discrepancies
in the position angle near $\rsma^{1/4}\sim 4$. In this region,
however, the ellipticity is so close to zero that the exact value of
the position angle has little meaning. The S\'ersic and 2dV fits for
M84 (given in Table \ref{sbfits}) yield a total luminosity of 7.2 and
6.5 $\times 10^{10} L_{\sun}$ respectively, and in the 2dV fit the
outer component carries 77\% of the total luminosity.

For M86, the comparison between our profiles and those previously
published is good throughout. In particular we note that the hump in
the surface brightness profile of M86 near $\rsma^{1/4}\sim 4$ is also
seen in the M86 profile of K09. This feature complicates the analytic
fitting process (Table \ref{sbfits}), yielding $\chi^2$ values
significantly worse than for any other galaxy in our sample. Our
S\'ersic fit differs dramatically from that of K09, but as K09 shows,
the range of radii chosen to fit the profile has a significant effect
on the fit parameters. K09 do not fit past $\rsma^{1/4} = 3.5$, and so
their fit excludes the hump. If we limit our fit to a similar range,
our fit parameters more closely match those of K09. The 2dV fit is
similarly poor, consisting of a small high surface brightness inner
component, and an outer component which contains 94\% of the
light. Given the actual shape of the profile, we do not consider this
2dV fit to be physically meaningful. The total luminosity of M86 is
9.3 and 9.2 $\times 10^{10} L_{\sun}$ under the S\'ersic and 2dV fits,
respectively.

We subtract the combined M86 and M84 models from the original image,
yielding the residual image shown in Figure
\ref{subtract_m84m86m89}. We note that in this case the mask displayed
on the final residuals is a subset of the mask that is used in the
fitting procedure. The bright galaxies south and east of M86 have all
been aggressively masked in the analysis, but are displayed in this
image for clarity. In the residual image, we immediately note the
pinwheel-like fins centered on M86, which are indicators of the boxy
isophotes noted by Peletier \etal (1990). These features are usually
represented as azimuthal A4 Fourier components, beyond a pure
elliptical model.  Since we do not include these higher-order Fourier
terms in our elliptical fits, these features show up as positive and
negative residuals in the subtracted image. We restrict our discussion of
the substructure around M86 to regions outside the area where these
terms dominate the residual light profile.

Around M86, we find several small radial streams, including ones to
the north (Region 1), south (Region 4), southeast (Region 5), and east
(Region 6). These small features are very unlikely to be fitting
artifacts, as they are coherent and consistent across many isophotes
of the galaxies.  We also find a few small plumes between M86 and M84
(Regions 2 and 3). We have verified that all these features can be seen in the
unsubtracted image, and are therefore not artifacts of the subtraction
process. It is tempting to link up features 1 and 4, and features 2 and 6,
into continuous streams that cross the face of M86, as would be
expected from tidal stripping of small orbiting satellites. The
luminosity of these features is small, typically $2-3 \times 10^7
L_{\sun}$ (Table \ref{m84m86tab}). A dogleg stream of light (Regions A
and B) also extends from the nearby interacting pair of galaxies NGC
4435/8. This nature of this feature, first identified by Malin (1994),
remains in doubt; optically it is bluer than other tidal streams in
Virgo (Rudick \etal 2010), and multiwavelength observations suggest
that it may in fact be a very unfortunate projection of galactic
cirrus across the galaxy pair (Cortese \etal 2010). 

We have also compared our deep optical image to the narrowband H$\alpha$
image of Kenney \etal (2008), who found a very complex system of
H$\alpha$ filaments connecting M86 with NGC 4438. Kenney \etal
proposed a collision between the ISM of the two galaxies as the source
of the filaments. If these galaxies are in collision, we see no strong
evidence of it in our imaging. We find no correlation between that
H$\alpha$ map and our deep imaging, and the few tidal features we see
between the galaxies (Regions 5 and 6) are small and narrow,
suggestive of small stripping of low velocity dispersion dwarf
galaxies. However this does not rule out an encounter -- the high
relative velocities of M86 and NGC 4438 ($\Delta v=1379$ km/s) would
suppress the formation of strong tidal features, but still drive the strong
response in the ISM/IGM giving rise to the ionized H$\alpha$ filaments
seen by Kenney \etal.

On smaller scales, Elmegreen \etal (2000) used optical imaging to
identify a number of dust streamers 10-20 kpc from the center of M86,
which they attributed to dust stripping from the dwarf elliptical VCC
882. Other evidence for dust stripping from galaxies orbiting M86
comes from far infrared ISOPHOT imaging (Stickel \etal 2003) which
revealed a number of infrared sources with spectra consistent with
cold dust emission lying within 35 kpc from M86. These features are
all close enough to M86 that they lie within the region where our
ellipse subtraction is confused by the $A4$ components in the galaxy
profiles, and so we have not tried to identify features here (although
we do see the optical dust lanes identified by Elmegreen \etal
2000). Nonetheless, the general inference that M86 is accreting and
stripping a population of dwarf galaxies is consistent with the
optical streamers we identify at larger radius.

As we did with M87, we have compared our residual image with the far
infrared maps to guard against confusion with galactic cirrus. The
infrared maps show significant dust contamination to the west of M84;
some of this cirrus can be seen as the (unlabeled) diffuse feature
seen in the upper right edge of our image. There is also infrared
emission associated with the dogleg plume near NGC 4435/8, one of the
arguments for its identification as galactic cirrus (Cortese \etal
2010). Other than these features, we see no clear correspondence
between the infrared dust emission and the diffuse streamers we have
identified near M84 and M86.

\begin{deluxetable}{ccccc}
\tabletypesize{\scriptsize}
\tablewidth{0pt}
\tablecaption{M84/M86 Substructure}
\tablehead{\colhead{Reg.} & \colhead{Name} & 
\colhead{Galaxy} & \colhead{$\mu_{V,\textrm{peak}}$} &
\colhead{L$_{\textrm{tot},V}$} \\
\colhead{\#} & \colhead{} & \colhead{} &
\colhead{[\magsec]} & \colhead{[$10^8$ $L_\odot$]} }
\startdata
1 & N Stream       &  M86     & 28.3 & 0.36 (0.09) \\
2 & N Middle Plume &  ?       & 27.6 & 0.17 (0.02) \\
3 & Middle Plume   &  ?       & 27.7 & 0.19 (0.03) \\
4 & S Stream       &  M86     & 28.1 & 0.18 (0.02) \\
5 & ESE Plume      &  M86     & 27.8 & 0.18 (0.03) \\
6 & E Stream       &  M86     & 28.2 & 0.19 (0.02) \\
\cline{2-5}
  & Total          &      & M86     &  0.91 -- 1.27 (0.10) \\
  & Total          &      & M84     &  0.00 -- 0.36 (0.04)
\enddata
\tablecomments{Uncertainties in the luminosities are given in
  parentheses. The total luminosities are given as ranges, depending
  on whether the luminosity of Regions 2 and 3 is given to M86 or M84.}
\label{m84m86tab}
\end{deluxetable}

\subsection{M89 (NGC 4552)}

The elliptical galaxy M89 has the lowest luminosity in our sample, and
resides one degree east of M87. M89's final mask eliminates all of the
bright stars and small galaxies in the field of view, and we also mask
the bright shell to the south and the strong ``jet'' feature to the
west, both identified first by Malin (1979). Since these features are
obvious substructures distinct from the smooth galaxy light, we want
to exclude them from contributing to the elliptical isophotal fit. We
again use a fixed center (of $\alpha=$12:35:39.8, $\delta=$+12:33:23.2
J2000) for our fit, and our best-fit surface brightness profile
reaches our limiting magnitude of \muv$ = 29$ at $\rsma$
=23\arcmin. Our best fit elliptical model is shown in Figure
\ref{allfits}, where we also compare our surface brightness profile
with that of K09, who combined original observations
with published data to construct a composite $V$-band profile for
M89. The comparison shows an excellent match in ellipticity and
position angle.

Our best-fit S\'{e}rsic and 2dV fits to the luminosity profile are
shown in Figure \ref{allfits} and the parameters are given in Table
\ref{sbfits}.  Our fitted S\'{e}rsic model has an extremely high index
of $n=14.6$, similar to the value of $n=13.9$ found by Caon \etal
(1993), but higher than the $n=9.2$ reported by K09. Again, however,
the radial range of the fit is important; in a fit with a radial range
that more closely matches ours, K09 derive a larger value of $n=13.75$
(see Figure 56 of K09). In our fits, the 2dV model yields a somewhat
better fit than the S\'ersic model, with the outer component
contributing 68\% of the total luminosity of $L_V=3.9 \times 10^{10}
L_{\sun}$. The S\'{e}rsic gives a higher total luminosity of $L_V=4.9
\times 10^{10} L_{\sun}$.

Subtracting our isophotal model from the raw image yields the residual
image shown in Figure \ref{subtract_m84m86m89}. As in the case of M86,
we see a strong signature of boxy A4 components in the inner parts of
the galaxy, and restrict our study of substructure to regions at
larger radius. A great deal of the area near M89 is covered in
substructure. Easily visible are the ``jet'' feature to the west
(Region 1) and the extremely bright shell to the south (Region 2). We
also find plumes to the northwest (Region 3), southwest (Region 4),
and southeast (Region 5) of M89. Fainter and further away are two
shells: one 9.3\arcmin\ (42 kpc) to the east (Region 6), and another
12.3\arcmin\ (57 kpc) to the northeast (Region 7).  As with the
previous galaxies, we have confirmed that all these features are
visible in the unsubtracted image as well.
Photometric measurements were made of each of the detected features
and compared with directly adjacent sky regions. Table \ref{m89tab}
gives the photometric properties for each feature.

\begin{deluxetable}{cccc}
\tabletypesize{\scriptsize}
\tablewidth{0pt}
\tablecaption{M89 Substructure}
\tablehead{\colhead{Region} & \colhead{Name} & 
\colhead{$\mu_{V,\textrm{peak}}$} & \colhead{L$_{\textrm{tot},V}$} \\ 
\colhead{\#} & \colhead{} &  \colhead{[\magsec]} &
\colhead{[$10^8$ $L_\odot$]} } 
\startdata
1 & W tail         &  25.7 & 2.56 (0.07) \\
2 & S shell        &  26.3 & 1.62 (0.05) \\
3 & NW plume       &  26.9 & 1.75 (0.07) \\
4 & SW Plume       &  27.8 & 1.09 (0.12) \\
5 & SE Plume       &  27.6 & 0.70 (0.05) \\
6 & E Shell        &  27.9 & 1.45 (0.23) \\
7 & NE Shell       &  28.2 & 0.57 (0.10) \\
\cline{2-4}
  & Total          &       & 9.74 (0.30)
\enddata
\tablecomments{Uncertainties in luminosities are given in parentheses.}
\label{m89tab}
\end{deluxetable}

The variety of tidal structure around M89 argues for a complicated
accretion history. The W Tail (Region 1) is peculiar in that it has a
high, relatively constant surface brightness across its length, and
may be a long tidal tail seen curving back on itself in
projection. Indeed there is some hint of a plume extending to the east
of M89 along the same axis as the W Tail, projecting across the outer
shells -- if real, this could be the extension of the W tail curving
back across the face of the galaxy. Unlike the shell system of M49,
M89's shells do not have the classic aligned and interleaved structure
expected from phase wrapping from a small accretion event (\eg Quinn
1984). Instead, the shells occur at a variety of position angles and
radii, suggesting material that has come into the galaxy with a range
of angular momenta. Such features could arise either from multiple
accretions, or via a major merger of two disk galaxies, where tidal
material can spatially wrap around the remnant without any co-alignment
of shells (\eg Hernquist \& Spergel 1992; Hibbard \& Mihos 1995). The
sharpness of the outer shells (Regions 6 and 7) also argues that, like
M49, M89 must not have experienced significant tidal stripping from
the Virgo cluster environment over the past Gyr or so, otherwise the
shells would have been disturbed or disrupted.

\section{Summary and Discussion}

In summary, we have used deep, wide-field surface photometry to study
the extended envelopes of five luminous Virgo ellipticals. We use the
IRAF ELLIPSE task to fit elliptical isophotes to the galaxies' surface
brightness profiles. Analytic fits to these isophotal models compare
well to fits published elsewhere in the literature. Subtracting the
isophotal models from the images, we identify a variety of low surface
brightness tidal features -- streams, plumes, and shells -- in the
outer envelopes of these elliptical galaxies, which we have quantified
in terms of their total luminosity and peak surface brightness. 
We find that M87 is characterized by an extended diffuse halo
with broad plumes and radial streams, but no sharp tidal shells or
loops. M86 shows a number of small, relatively high surface streams,
while M84 shows no evidence for significant substructure beyond its
smooth elliptical isophotes.  In contrast, both M49 and M89
show complex (and in M49, hitherto undiscovered) systems
of distinct shells and other tidal features. 

\begin{figure*}[]
\includegraphics[width=6.75in]{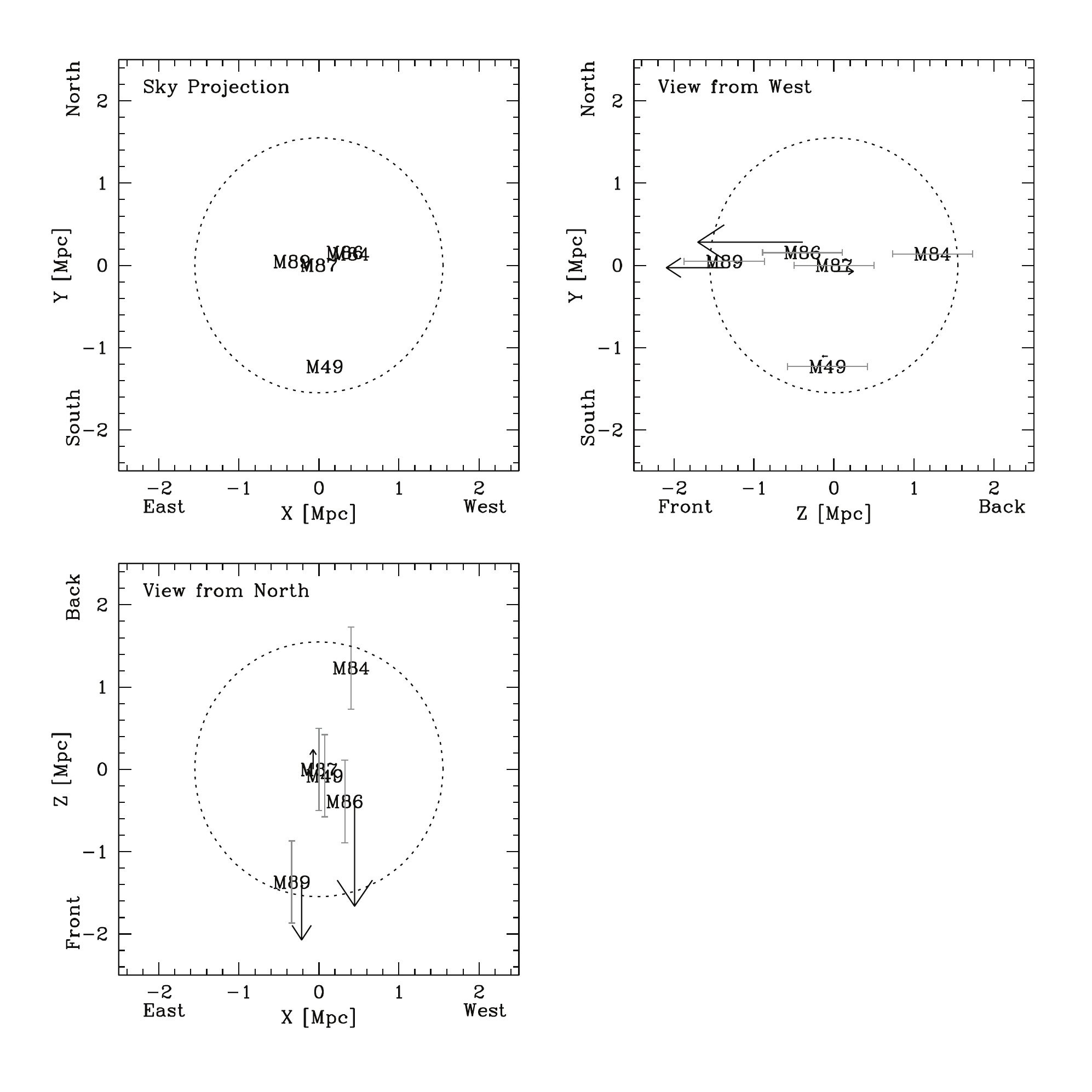}
\caption{The three dimensional geometry of the Virgo cluster, using
  line of sight SBF distances from Mei \etal (2007). In each panel,
  the dotted line shows the virial radius ($r_{200}$) for Virgo from
  McLaughlin (1999). Arrows show the line of sight velocity of each
  galaxy, with the length of the arrow equal to the distance traveled
  in 1 Gyr.}
\label{virgogeom}
\end{figure*}

The variety of structure we see in these Virgo ellipticals may well
reflect differences in their history of accretion of smaller galaxies,
their accretion history into the Virgo Cluster itself, or a
combination of the two. To begin our discussion of environmental
influences, we show in Figure \ref{virgogeom} three orthogonal
projections of the three dimensional positions of our sample galaxies
in the Virgo cluster. We use the distances to each galaxy derived by
Mei \etal (2007) using the surface brightness fluctuation
technique. We adopt this dataset largely because it has consistently
derived distances for all of our galaxies, with very low internal
errors. However, there are systematic uncertainties in the absolute
distances; other techniques yield somewhat different results. Most
notably, the position of M86 is quite uncertain -- while the Mei \etal
distances place M86 in the cluster core, in front of M84, distance
determinations derived from planetary nebula put both M86 and M84 at
the same distance, 1 Mpc behind M87 (Jacoby \etal 1990). Finally, Figure
\ref{virgogeom} also plots the line of sight velocity vectors of each
galaxy, after subtracting out a mean Virgo velocity of 1064 km/s
(Binggeli 1999), and the estimated r200 (1.55 Mpc) for Virgo,
determined by McLaughlin (1999).

Without transverse velocities for these galaxies, a full dynamical
interpretation is impossible; nonetheless, we can glean useful
information from this plot. As Virgo's central galaxy, it is no
surprise that M87 has a low line of sight motion ($v_{\rm rel} = 243$
km/s) with respect to the cluster as a whole. Sitting $\sim$ 1 Mpc
behind the cluster core, M84's low velocity ($-4$ km/s) suggests it is
either near apocenter on a low angular momentum orbit -- likely having
passed near the cluster center a few Gyr ago -- or moving on a more
tangential orbit which keeps it out of the cluster center. As M49 is
not projected onto the cluster core, its low velocity ($-67$ km/s)
places little constraint on its orbit. However, the presence of an
X-ray bowshock to the north of M49 argues that the galaxy is falling
into Virgo's hot intracluster medium from the south (Irwin \& Sarazin
1996). Only M86 and M89 have significant line of sight motion with
respect to the cluster center ($-1308$ km/s and $-724$ km/s,
respectively). M86 is either moving at high speed through the core, or
just about to enter it, depending on the adopted distance, while M89
either passed through the core $\sim$ 1-2 Gyr ago (if it lacks
significant transverse velocity) or is on a more tangential orbit that
keeps it further from the core. 

\begin{deluxetable}{cccccc}
\tabletypesize{\scriptsize}
\tablewidth{0pt}
\tablecaption{Galaxy Properties}
\tablehead{\colhead{Galaxy} & \colhead{$R_p$\tablenotemark{a}} &  \colhead{$r_{3d}$\tablenotemark{b}} & \colhead {$v_{\rm rel}$\tablenotemark{c}} & \colhead{\Lsub} &  \colhead{\fsub} \\
\colhead{ }  & \colhead{[Mpc]} & \colhead{[Mpc]} & \colhead{[km/s]} & \colhead{[$10^8 L_\odot$]}  & \colhead{[\%]} } 
\startdata
M49 & 1.23 & 1.32 &   -67 & 7.29 (0.25) & 0.47 (0.02) \\
M87 & $\equiv 0$    & $\equiv 0$    &   243 & 4.42 (0.46) & 0.39 (0.04) \\
M86 & 0.35 & 0.54 & -1308 & 0.91 - 1.27 & 0.10 - 0.14 \\
M84 & 0.42 & 1.32 &    -4 & 0.00 - 0.36 & 0.00 - 0.05 \\
M89 & 0.33 & 1.41 &  -724 & 9.74 (0.30) & 1.99 (0.06) \\
\enddata
\tablenotetext{a}{Projected clustercentric distance, defined relative
  to M87}
\tablenotetext{b}{Three dimensional clustercentric distance}
\tablenotetext{c}{Velocity relative to cluster velocity of +1064 km/s
  (Binggeli \etal 1999)}
\label{galprop}
\end{deluxetable}

We first examine the environmental question by looking to see if the
amount of diffuse substructure in the galaxies correlates with their
distance from the center of Virgo (taken to be the position of
M87). We quantify the amount of substructure in two ways: the total
luminosity in the features (\Lsub), and their fractional luminosity
(\fsub, measured with respect to the total galaxy light). In the
latter case, the galaxy luminosities are calculated analytically from
our S\'ersic surface brightness fits. We then calculate each galaxy's
distance from the center of the Virgo Cluster using a combination of
their projected distance from M87 on the sky and their line-of-sight
distance from Mei \etal (2007). These quantities are given in Table
\ref{galprop}.

\begin{figure*}[]
\includegraphics[width=6.75in]{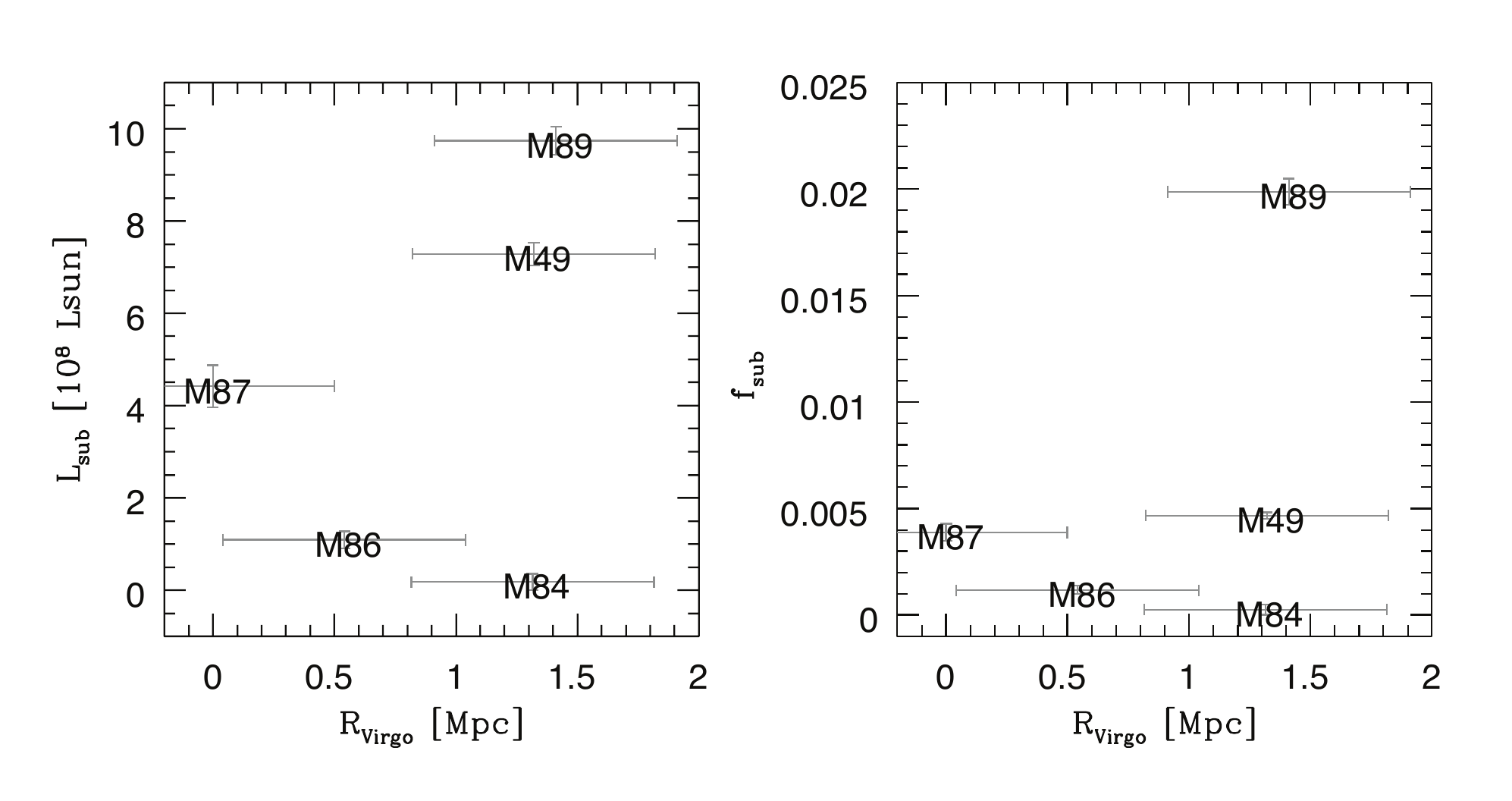}
\caption{Substructure luminosity (left) and luminosity fraction
  (right) as a function of clustercentric distance.} 
\label{environ}
\end{figure*}

For these five galaxies, we see no obvious correlation between the
amount of substructure and cluster-centric distance, as shown in
Figure \ref{environ}. M87, M49, and M89 have significantly more
substructure luminosity than the others, although in the case of M87
and M49, the substructures contribute much less to the total
luminosity of the galaxies. Neither M84 nor M86 appear to have
appreciable substructure, either in total or fractional luminosity. A
model where substructure survival depends simply on cluster-centric
distance is clearly too simplistic to explain the features seen in
these Virgo ellipticals.

However, M87 is not simply {\it deep} in the cluster potential, it
lives {\it at the center} of the cluster. In this privileged position,
it experiences a constant rain of smaller satellite galaxies falling
in at high velocities, which will be tidally stripped and leave behind
long streamers such as those seen extending to the NW of M87. Unlike
the long-lived structures around isolated galaxies, streamers inside
an active cluster environment are dispersed rapidly via interactions
with other cluster members, typically within a few crossing times
(Rudick \etal 2009). In this position, M87 has both high creation
rates and high destruction rates for its diffuse substructure, so that
the window of opportunity for observing cold streams around M87 will
be short.

While the amount of substructure does not seem to correlate with
cluster-centric radius, there are interesting patterns in the
morphological properties of the diffuse light in these
ellipticals. M49 and M89 both have complex shell structures out to
large radius, where the material would be loosely bound to the host
galaxy. Any strong tidal forces that would occur during a galaxy's
passage through the cluster would likely be sufficient to strip or at
least significantly perturb the shells (see, \eg Mihos 2004). The
sharpness of the shells and the long dynamical timescales in the
outskirts of these galaxies ($\sim$ 0.5 Gyr) argue that these galaxies
have not passed through the dense Virgo Core in the recent past.  In
contrast, we see little evidence for sharp shell-like structures in
M87, M86, or M84. Living at the center of Virgo, M87 experiences
repeated encounters with cluster galaxies, and any dynamically
delicate feature will quickly be destroyed.  In the radial streams
visible to the northwest of M87, we are likely seeing very recent
stripping of galaxies falling in on radial orbits.

The nature of the streams observed in M87 and M86 is qualitatively
different as well. M87's NW and WNW streams can be traced out to
$\sim$ 1 degree (275 kpc) from the center of M87, while M86's streams
are much smaller. The longest, the N Stream, extends 25\arcmin\ from
M86, while the others are much smaller in length (only a few
arcminutes in size) and closer to M86. Moving at such a high velocity
near or through the cluster core, M86 simply may not be able to hang
on to extended debris. The cluster tidal field may easily strip these
streams, or they may lack coherency simply because of the motion of
M86. A coherent stream at 300 kpc would have an orbital period of a
few Gyr, a timescale over which M86 would have moved a few Mpc through
the cluster. The combination of tidal effects and M86's high speed
motion through the cluster would make it difficult for M86 to retain
highly extended streamers.

Substructure is not the full story of the diffuse light in Virgo
ellipticals, of course. Because these features are likely to be
short-lived in a dynamically complex environment like a galaxy
cluster, they will mix into a more diffuse, extended envelope. Whether
or not this envelope is a dynamically or structurally distinct entity
from the galaxy itself is a subject of considerable debate. Gonzalez
\etal (2005) argue that in clusters with clear BCGs, the intracluster
light settles into a structurally distinct component from the galaxian
light of the BCG, and model the two components using distinct $r^{1/4}$
profiles. They find that, in their sample of BCGs, the outer
profile typically contains $\sim$ 80-90\% of the total luminosity, and
has an effective radius 10--40 times larger than the inner component.
If we consider M87 in this way, we find a smaller fraction of 
light in the extended envelope (50\%) than typical in the Gonzalez
\etal sample. However, Virgo is somewhat different from those BCG
clusters, in that it has a number of comparably bright galaxies
(M87 at the center; M86 projected 0.5\degr\ away; and M49 2\degr\ to the
south). With multiple bright galaxies and both spatial and kinematic
substructure, Virgo may represent a dynamically less-evolved
progenitor to the Gonzalez \etal clusters, in which case the lower fraction
of light in the envelope may be a signature of an still-developing ICL
(\eg Rudick \etal 2006). This inference echoes that of K09, who argue 
based on the systematics of M87's S\'ersic fit that the galaxy is at
best only a {\it weak} cD galaxy.

In this context, it is also interesting to look at M49. M49 is the
dominant galaxy in the Virgo Southern Extension (VSE), and somewhat
more luminous than M87. If clusters build through the hierarchical
accretion of sub-clumps, the sub-clumps themselves may have their own
diffuse light component, as suggested in the simulations by Rudick
\etal (2006). If the VSE is an evolved group now being accreted into
the Virgo cluster, it may have a structure similar to (albeit smaller
than) that of the BCG clusters of Gonzalez \etal (2005). And indeed,
the 2dV fit for M49 shows a higher fraction of light (76\%) in the
outer component than does M87, but with a much smaller characteristic
scale ($r_{e,out}$(M49) = 20\% $r_{e,out}$(M87)).

While these results are consistent with a scenario of a gradual,
on-going buildup of ICL in Virgo, there are several important
caveats. First, the 2dV profiles are not clearly superior to regular
S\'ersic fits, and the justification for dividing luminosity into an
inner and outer component based on these fits is not strong. Even the
choice of functional form for the division is the source of some
debate -- Seigar \etal (2007) argue that an inner S\'ersic + outer
exponential fit is a better description of the light profile for cD
galaxies. However, given the fact that the additional free parameter
of the 2dV fit over the S\'ersic fit did not result in significantly
better profile fits for our galaxies, we do not pursue these higher
order fits here.

Given our results, we paint a plausible picture of the dynamics of the
galaxies within the Virgo Cluster. Sitting at the center of the
cluster, M87 experiences a rain of smaller galaxies which are being
tidally stripped by the cluster potential, leading to the long diffuse
streams seen to the NW of M87. Due to encounters with other galaxies
in the cluster core, the lifetime of these streams is short and they
mix away to continually build M87's extended envelope and Virgo's
intracluster light. The combination of M86's high velocity and its
passage through the cluster core makes it difficult to develop or
retain very extended streams; instead, it possesses a system of small
tidal streams much closer to the galaxy than seen in M87. The lack of
tidal structures around M84 may be due to a possible recent passage
through the cluster core, or M84 may simply have not experienced much
recent accretion.

In contrast, both M49 and M89 display extended ($r \sim 50-100$ kpc)
shell systems, arguing that they have not experienced the strong tidal
forces of the Virgo cluster core in the past Gyr or so. As the most
luminous galaxy in the Virgo Southern Extension, it may be the
dominant galaxy of group falling into the Virgo cluster for the first
time, and its shell system reflects its own accretion of a smaller
satellite on a radial orbit. As M49 falls into Virgo, its shell system
will be disrupted and incorporated into the general ICL of the
cluster. The complexity of M89's shell system argues not for an
individual accretion event, but rather for multiple satellite
accretions or a major merger in its past. Its system of tails, plumes,
and shells will likely also be dispersed into Virgo's extended ICL as
the galaxy orbits within the cluster environment.

Ultimately, however, morphology alone contains only limited
information. A better understanding of these features and how they
relate to the dynamical history of Virgo and its galaxies would come
through studies of their kinematics and stellar populations. Kinematic
information can come from identification and follow-up spectroscopy of
planetary nebulae (\eg Arnaboldi \etal 2004, Doherty \etal 2009) and
globular clusters (\eg C\^ot\'e \etal 2003; Hwang \etal 2008;
Lee \etal 2010) associated with the streams, while stellar populations
can be studied using deep multiband surface photometry (Rudick \etal
2010) or {\it Hubble Space Telescope} imaging
of resolved stars in the streams (\eg Williams \etal 2007). Such
studies would give an integrated picture of the accretion and
stripping processes at work in the Virgo Cluster.

\acknowledgments

We thank Charley Knox for his tireless work on the mechanical,
optical, and electronic support of the Burrell Schmidt.  Over the
course of this project, JCM has been supported by the NSF through
grants AST-9876143, ASTR-0607526, and AST-0707793, as well as by
Research Corporation through a Cottrell Scholarship. CR was supported
for part of this work by the Jason J. Nassau Graduate
Fellowship fund.

{\it Facility:} \facility{CWRU:Schmidt}

\end{document}